\documentclass[%
 aip,
 amsmath,amssymb,
 reprint,%
]{revtex4-1}

\usepackage{graphicx}% Include figure files
\usepackage{dcolumn}% Align table columns on decimal point
\usepackage{bm}% bold math
\usepackage[utf8]{inputenc}
\usepackage[T1]{fontenc}
\usepackage{mathptmx}
\usepackage{graphics}

\usepackage{physics}
\begin{document}

\preprint{AIP/123-QED}

\title[The interaction of twisted Laguerre-Gaussian light with a GaAs photocathode to investigate photogenerated polarized electrons]{The interaction of twisted Laguerre-Gaussian light with a GaAs photocathode to investigate photogenerated polarized electrons}
% Force line breaks with \\

\author{L. A. Sordillo}
\email{lsordillo@ccny.cuny.edu}
\author{S. Mamani}
\author{M. Sharonov}
\author{R. R. Alfano}
\affiliation{ Institute for Ultrafast Spectroscopy and Lasers, Department of Physics at The City College of the City University of New York, 160 Convent Avenue, New York, NY 10031, USA.\\}
\affiliation{The Grove School of Engineering, Department of Electrical Engineering at The City College of the City University of New York, 160 Convent Avenue, New York, NY 10031, USA. 
}

\date{Submitted: October 25th, 2018; Accepted: January 15th, 2019}% It is always \today, today,

\begin{abstract}
The interaction of Laguerre-Gauusian light at select wavelengths from 690 to 810 nm with a p-type gallium arsenide (GaAs) photocathode photonic device and the production of photogenerated electrons was investigated.  Spin angular momentum (SAM) and orbital angular momentum (OAM) were generated using linear or circular polarization and q-plates. The degree of polarization P from the photogenerated electron signal was measured. At an excitation of 695 nm ($E=1.78$ $eV$), $P_{OAM}=~2.1 \%$, $P_{SAM}=~1.3 \%$ and $P_{SAM, OAM (q=1)}= ~3.4\%$, whereas, at 800 nm ($E=1.55$ $eV$), near bandgap, $P_{OAM}=~-14.7\%$, $P_{SAM}=~-2.6\%$  and $P_{SAM,OAM (q=1)} = ~4.0\%$. 
\end{abstract}

\maketitle

Ever since the first report of spin-polarized electrons in a semiconductor from circular polarized light in the 1960s, there has been considerable interest among high energy and condensed matter physicists on using circular polarized light to generate spin-polarized electrons in semiconductors [1, 2]. Circularly polarized light beams carrying spin angular momentum (SAM) have been used to orient the electron spins in semiconductors, causing formation of spin-polarized photogenerated electrons. In 1968, Lampel showed that spin-polarized electrons in semiconductor silicon (Si) can be produced by optically pumping with circular polarized light [1]. A year later, Parsons reported on band-to-band optical pumping of electrons in p-type GaSb [2]. 

Gallium arsenide (GaAs) has several advantages over Si and other semiconductors. Ekimov and Safarov investigated the optical orientation of free carriers in n- and p-type $Ga_xAl_{1-x{\pm}}$ crystals (same band structure as GaAs) pumped by circular polarized light from a He-Ne laser with $E=1.96$ $eV$ [3]. Zakharchenya et al showed the effect of optical orientation of electron spins in a GaAs crystal using a wide range of excitation energies from $1.5$ to 2.2 $eV$, and Pierce et al detected spin-oriented electrons from GaAs using a Mott detector and circular polarized light at energies from 1.5 to 3.6 $eV$ [4, 5]. Seymour and Alfano measured directly spin relaxation times in p-GaAs (Zn doped, NA= 7$\times$$10^{17}$/cm$^3$) crystal using time-resolved spectroscopy [6- 9]. Spintronic devices rely on the control or manipulation of electron spins in semiconductors and have been proposed for computation, storage and information applications. Recently, due to its n-dimensionality, orbital angular momentum (OAM), a quantum property of light, is gaining attention as a potential way to hold more information.

This Letter focuses on the use of SAM and OAM light from a tunable 100 femtosecond Ti:sapphire laser at wavelengths from 690 to 810 nm to investigate photogenerated electrons from a compact photonic device. The photonic device, with a p-type GaAs (cesium) photocathode (at a concentration of approximately $10^{18}$/cm$^3$) and lock-in amplifier, was utilized to visualize the degree of spin orientation of the optically pumped electrons. In fact, the GaAs photocathode photonic device has a dual role, acting as a sample and as a detector. The photonic device was pumped using three select types of light beams carrying 1) SAM (circular polarization), 2) OAM $l=\pm1, +2$ (azimuthal polarization generated from linear polarized light and q-plates $q=\pm1/2, +1$) and 3) SAM with OAM $l=\pm1, +2$ (light generated from circular polarization and $q=\pm1/2, +1$), producing a photogenerated electron signal. The degree of polarization of the photogenerated electron signal was calculated for $P_{OAM}$, $P_{SAM}$ and $P_{SAM, OAM}$ in terms of the quantum excitation energies from $E= 1.53$ to 1.78 $eV$. The use of SAM and OAM to optically pump electrons is described for electrons in a semiconductor GaAs material. Though several recent studies have emerged on the absorption of OAM by atoms and molecules [10-15], limited studies have been reported on the use of OAM to optically pump electrons in GaAs [16-17]. The following subsections highlight the important features of SAM and OAM in GaAs.

\section{\label{sec:level1}Spin Angular Momentum}
Spin-polarized electrons in GaAs are typically produced using photons carrying SAM in the form of right $\ket{R}$ or $\ket{L}$ (helicity $\sigma^+$ or $\sigma^-$, respectively) circular polarized light. The SAM light optically pumps the electron population  from the valence band to the conduction band in select transitions ($\Gamma_{8h}$ $\rightarrow$, $\Gamma_{6}$, $\Gamma_{8l}$ $\rightarrow$ $\Gamma_{6}$ or $\Gamma_{7}$ $\rightarrow$ $\Gamma_{6}$), producing spin-polarized electrons in the conduction band. Fig. 1 shows the selection rules using SAM and the energy bands for a GaAs semiconductor in k-space with respect to $\Gamma$ at $k=0$. 
\begin{figure}[h]
\includegraphics [width=80mm, height=45mm] {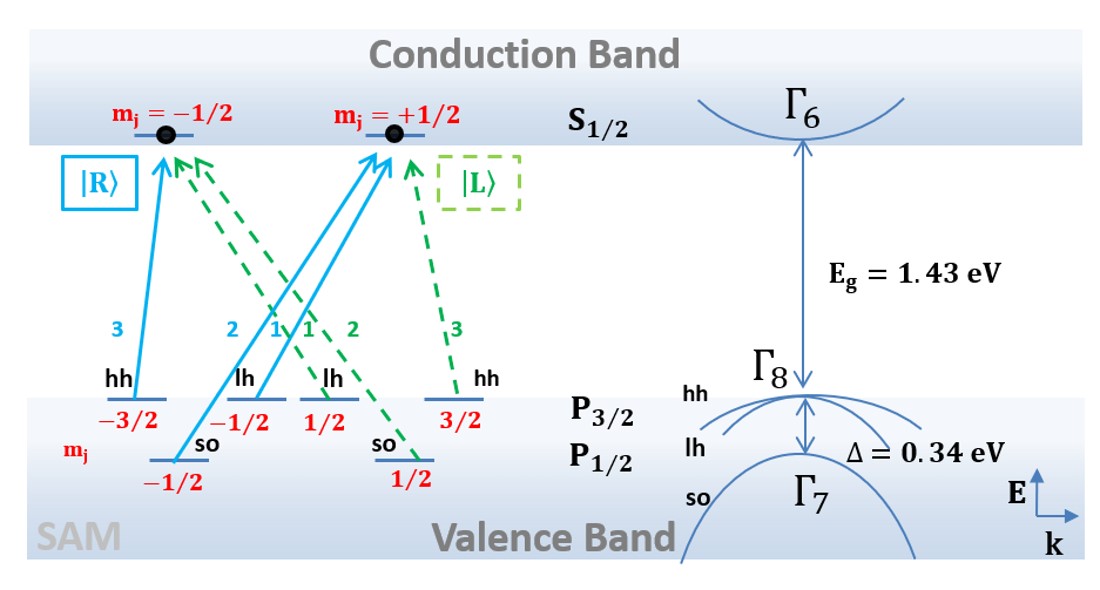}
\caption{Selection rules and the energy bands for GaAs in k-space with respect to the center of the Brillouin zone $(\Gamma)$ at $\ k=0$.} 
\end{figure}
The valence band (p-symmetry) is divided into four-fold degenerate $ P_\frac{3}{2}$ states at $\Gamma_{8}$; a heavy hole (hh) at the $\Gamma_{8h}$ band and light hole (lh) at the $\Gamma_{8l}$ band. In the presence of spin-orbit interaction, the split-off (so) valence band at $\Gamma_{7}$ is separated from $\Gamma_{8}$ by an energy difference of $\Delta$= .34 $eV$ and consist of two-fold degenerate $ P_\frac{1}{2}$ states. The s-symmetry conduction band, with two-fold degenerate $ S_\frac{1}{2}$, is located at $\Gamma_{6}$. At room temperature (300 K), the upper valence bands at $\Gamma_{8}$ and the lowest conduction band at $\Gamma_{6}$ are separated by an energy gap $E_g$= 1.43 $eV$ $(E_g= 1.52$ $eV$ at 0 K).

By changing the wavelength of the incident laser light, select transitions from the sublevels can be obtained. For right circular polarized SAM light near bandgap, only transitions from the $m_j$= -1/2 (valence band) to the $m_j$= +1/2 (conduction band) and $m_j$= -3/2 (valence band) to $m_j$= -1/2 (conduction band) can occur. The transition probabilities are determined by calculating the matrix element of the transition for the SAM light [18, 19]. At $h\nu$ less than 1.76 eV, $\bra{x, \sigma}$$\ket{S,s}$= $u_c(r)$ $\chi_c(\sigma)$, where $\sigma$= $\downarrow, \uparrow$ and r= $\mid$x$\mid$ is the modulus of x. Spherical harmonics $Y_{lm}$ (angular part of the wave function) and $u_{\nu,c}(r)$ (radial part) describe the heavy hole and light hole transitions and can be written as $\ket{J,m_j}$ in terms of the total angular momentum (J) and projection quantum number ($m_j$) in equations (1)-(4)
\begin{equation}
\bra{x,\sigma}\ket{J=\tfrac{3}{2}, m_j=\tfrac{3}{2}}= Y_{11}(\hat{x})\chi\uparrow(\sigma)u_v(r),
\end{equation} 
\begin{equation}
\bra{x,\sigma}\ket{\tfrac{3}{2}, \tfrac{1}{2}}= \tfrac{1}{\sqrt3}[\sqrt2Y_{10}(\hat{x})\chi\uparrow(\sigma) + Y_{11}(\hat{x})\chi\downarrow(\sigma)]u_v(r),   
\end{equation} 
\begin{equation}
\bra{x,\sigma}\ket{\tfrac{3}{2}, -\tfrac{1}{2}}= \tfrac{1}{\sqrt3}[\sqrt2Y_{10}(\hat{x})\chi\downarrow(\sigma) +Y_{1-1}(\hat{x})\chi\uparrow(\sigma)]u_v(r),   
\end{equation}
\begin{equation}
\bra{x,\sigma}\ket{\tfrac{3}{2}, -\tfrac{3}{2}}= Y_{1-1}(\hat{x})\chi\downarrow(\sigma)u_v(r).
\end{equation}
As the energy is increased, transitions from the split-off valence band are possible. Equations (5) and (6) describe the functions for the split-off hole, 
\begin{equation}
\bra{x,\sigma}\ket{\tfrac{1}{2}, \tfrac{1}{2}}= \tfrac{1}{\sqrt3}[\sqrt2Y_{11}(\hat{x})\chi\downarrow(\sigma) - Y_{10}(\hat{x})\chi\uparrow(\sigma)]u_v(r),
\end{equation}
\begin{equation}
\begin {split}
\bra{x,\sigma}\ket{\tfrac{1}{2}, -\tfrac{1}{2}}= \tfrac{1}{\sqrt3}[\sqrt2Y_{10}(\hat{x})\chi\downarrow(\sigma) \\
-\sqrt{2}Y_{1-1}(\hat{x})\chi\uparrow(\sigma)]u_v(r).
\end{split}
\end{equation}
The ratio between heavy hole and light hole transitions for $\hat{\sigma}_{\pm}$	$\propto Y_{1\pm1}$ can be calculated using equation (7)
\begin{equation}
\bra{S,s}\hat{\sigma}_{\pm}\ket{J,j}
\end{equation}
where S is the total spin and s is the spin projection. This leads to the ratio of transition strengths in equation (8)
\begin{equation}
\abs{{\frac{\bra{S=\tfrac{1}{2}, s=\downarrow}\hat{\sigma}_+\ket{J=\tfrac{3}{2}, j=-\frac{3}{2}}}{\bra{S=\tfrac{1}{2}, s=\downarrow}\hat{\sigma}_+\ket{J=\tfrac{3}{2}, j=-\tfrac{1}{2}}}}}^2=3
\end{equation}
 The strengths of the probabilities are marked as 1, 2 and 3 in Fig.1, where 3 is the greatest relative transition probability. The selection rules for light carrying OAM differ from the selection rules for SAM beams.

The spin polarization of the photogenerated GaAs electrons can be measured using equation (9)
\begin{equation}
P=\frac{N\uparrow-N\downarrow}{N\uparrow+N\downarrow},
\end{equation}
where $N\uparrow$ represents the number of electrons with spin parallel (up) and $N\downarrow$ the number of electrons with spin anti-parallel (down). Due to the degeneracy in energy of the hh and lh bands at the $\Gamma$-point, the strength of the interband matrix element of the hh transition is three times stronger than the lh transition. Thus, the theoretical spin polarization maximum P is $P_{theoretical}= {50\%}$. However, the degree of the circular polarization of the photoluminescence is $25\%$ as ${\tau_{pe}\ \gg\tau_s}$, due to a lack of electron spin relaxation during the photoemission escape time $\tau_{pe}$ 
\begin{equation}
P=P_{theoretical}\frac{\tau_s}{\tau_{pe}+\tau_s},
\end{equation}
where $\tau_s$ is the spin relaxation time from the conduction band [1, 5]. The degree of polarization as a function of time $P_s(t)$ can be written as equation (11):
\begin{equation}
P_s(t)=\frac{N\uparrow-N\downarrow}{N\uparrow+N\downarrow} \exp^\frac{-t}{(\tau_s/2)},
\end{equation}
where the spin alignment decays with an exponential time $\frac{1}{2}$$\tau_s$ [6-8, 18]. Seymour and Alfano found that the initial spin alignment for p-GaAs was 50 $\pm$ {10\%}, $\frac{1}{2}$$\tau_s$ was 44 $\pm$ 17 ps and $\tau_{pe}$ was 196 $\pm 12$ ps  [6]. Nowadays, the degeneracy between hh and lh at the valence band maximum is resolved by using GaAs/GaAsP and InGaAs-AlGaAs strained-layer superlattice and quantum well structures, for example, with maximum polarization of {~92\%} [20]. 
\section{\label{sec:level1}Orbital Angular Momentum}
Recently, optical pumping techniques, utilizing multiphoton absorption or Laguerre-Gaussian modes with different amounts of orbital angular momentum (OAM), have been used to excite electrons in GaAs. These studies resulted in electron polarizations of $\sim$$42\%$ and {3\%}, respectively [16, 17]. Unlike SAM, which has two states (spin up or spin down), OAM has countless states. Thus, OAM may be utilized in areas including quantum information technology, optical tweezing and optical information theory. OAM is described quantum mechanically as discrete intervals of $l\hbar$ and by the helical (twisted) phase factor $e^{il\varphi}$ (where $l$ is the topological charge and $\varphi$ is the azimuthal direction) and possesses zero momentum at the center of the vortex [21-23]. Fig. 2 (adapted from Clayburn et al) shows selection rules for light carrying OAM with $l=+1$ and $+2$ [17]. For OAM $l=+1$, there are several transition paths for the electron. These transitions have a net angular momentum of either 0 or +2$\hbar$, due to conservation of momentum. For $l=+2$ the transitions are primarily from the negative spins in the both p-symmetry valence bands to the conduction band [17, 24, 25]. There are numerous methods which can be utilized to generate these beams (SLM, optical vortex retarder and q-plates) [26]. The q-plate is a patterned birefringent liquid crystal plate with semi-integer or integer topological charge q [27-33]. The polarization transformation from the q-plates in the xy plane is given by the Jones matrix in equation (12)
\begin{equation}
M_q=
\begin{pmatrix}
\cos{2q}\theta & \sin{2q}\theta \\
\sin{2q}\theta & -\cos{2q}\theta
\end{pmatrix}
\end{equation}
where the azimuthal angle $\theta=\arctan{\frac{y}{x}}$ .When $\ket{R}$ or $\ket{L}$ from a quarter wave plate (QWP) falls upon a q-plate, the output beam will be in an entangled optical state of polarization and OAM.
\begin{figure}[h]
\includegraphics [width=80mm, height=45mm] {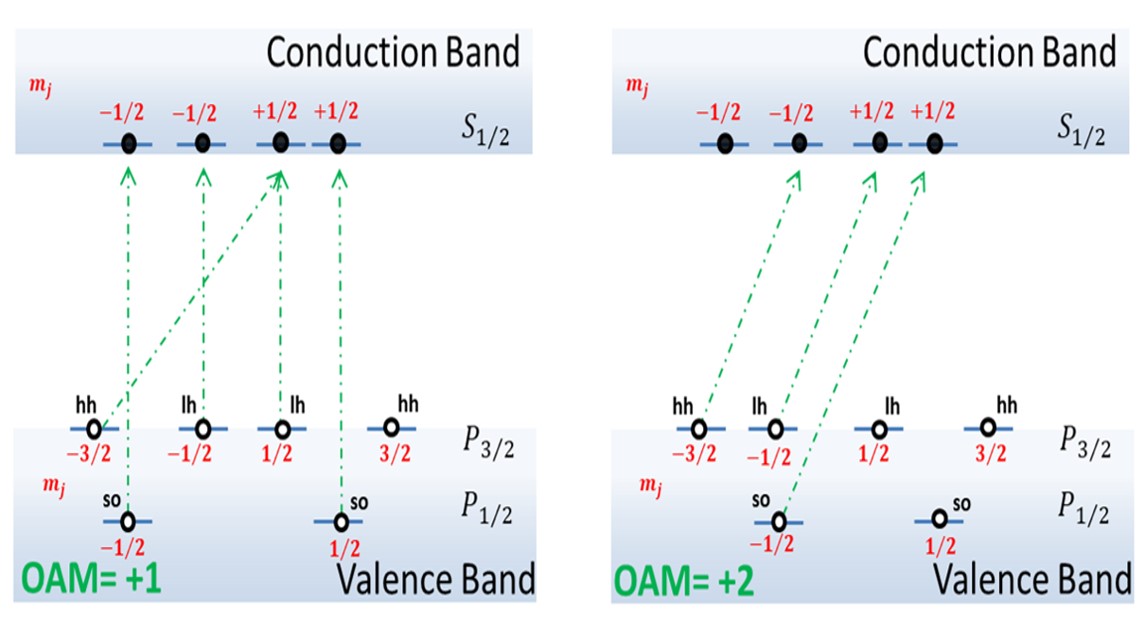}
\caption{Selection rules for light carrying OAM= +1 and +2 (adapted from Clayburn et al [17]).}
\end{figure}

Linear polarized light from a Coherent Ti:sapphire laser at 100 mW of power was used to create light with entangled SAM and OAM. The wavelength of the light was selected in increments of 5 nm from 690 to 810 nm, which corresponds to the energy bands of GaAs. Right and left circular polarized light carrying SAM was created from linear polarized laser light and a rotating quarter wave plate (QWP), while twisted photons carrying OAM were obtained using q-plates of different topological charge q= $\pm1/2$ and +1. Fig. 3(a) shows the optical setup with linear polarized light and a QWP, resulting in either left or right circular polarized light (SAM). Liquid crystal q-plates with a signal generator and topological charge q=-1/2, +1/2 or +1 were used to generate OAM light (Fig. 3(b)). SAM and OAM light was created using a q-plate and QWP and is shown in Fig 3(c). The plates were kindly provided by Professor Lorenzo Marrucci of the Universita' di Napoli Federico II [33]. An important feature of this study is the ability to produce either radial or azimuthal polarization.
\begin{figure}[h]
\includegraphics [width=80mm, height=55mm] {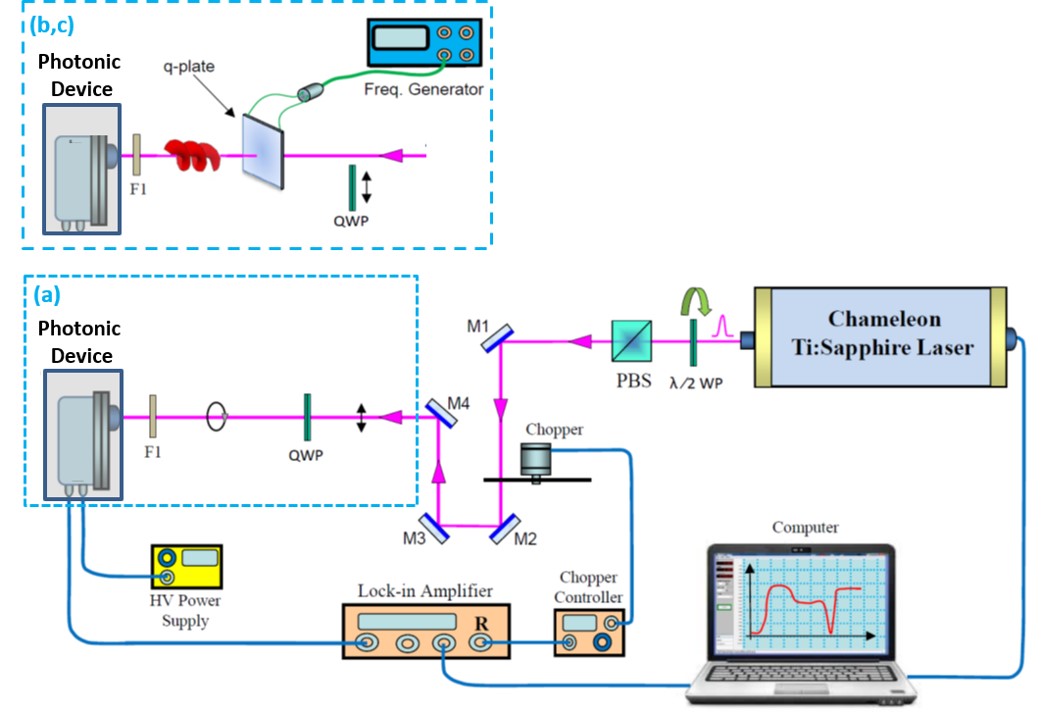}
\caption{The optical setup includes a Ti:sapphire laser, a GaAs photonic device and a lock-in amplifier. The optical setup was adjusted for three cases: (a) a QWP to produce SAM light, (b) a q-plate to produce OAM light and (c) a q-plate and QWP to produce SAM and OAM light. }
\end{figure}

The optical state of the beam before and after passing through a q-plate can be described using ket notation: $\ket{polarization, OAM(l)}$. When light carrying horizontal $\ket{H}$ linear polarization $\ket{LP}$ (a superposition of $\ket{L}$ and $\ket{R}$) passes through a q-plate with topological charge q, the resultant light beam has azimuthal polarization and is described by equation (13): 
\begin{equation}
\ket{H, 0} \rightarrow \frac {(\ket{R, +2q} +\ket{L,-2q})}{\sqrt{2}},
\end{equation}
where $\ket{L}$ and $\ket{R}$ represent left and right circular polarized light, respectively, and $\ket{0}$ is a superposition of +1 and -1 OAM. Thus, for q=+1/2, the $\ket{H}$ light beam is described as shown in equation (14)
\begin{equation}
\ket{H, 0} \rightarrow \frac {(\ket{R, +1} +\ket{L,-1})}{\sqrt{2}},
\end{equation}
Similarly, for q=+1, the output light beam will be described by equation (15)
\begin{equation}
\ket{H, 0}\rightarrow \frac {(\ket{R, +2} +\ket{L,-2})}{\sqrt{2}}  
\end{equation}
However, when light carrying $\ket{L}$ or $\ket{R}$ circular polarization emerges from a q-plate, for example from q=+1/2, the resultant beam is described by equations (16) and (17)
\begin{equation}
\ket{R, 0}\rightarrow\ket{L, -1}
\end{equation}
\begin{equation}
\ket{L, 0}\rightarrow\ket{R, +1}
\end{equation}

In each scenario, the photogenerated electron signals were recorded using a GaAs photocathode photonic device. The photonic device (Hamamatsu model number R636-10) is made of photocathode material p-type GaAs with a spectral response from 185 to 930 nm. The GaAs surface was treated with cesium (Cs) in order to obtain a negative electron affinity. A lock-in amplifier with a reference chopper at a frequency of 0.9 kHz was used. The electron signal intensities from optical pumping (incident light) versus time in seconds was recorded and displayed using a computer program.
\begin{figure}[h]
\includegraphics [width=80mm, height=55mm] {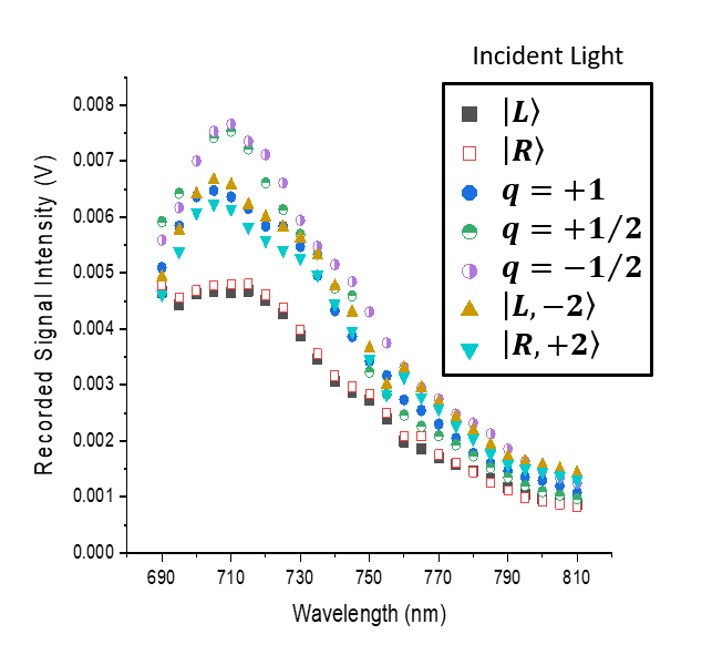}
\caption{Intensity of the recorded photogenerated electron signals based on optical pumping with SAM and OAM incident light (labeled in terms of the input beams) versus wavelength (nm)}
\end{figure}
Fig. 4 summarizes the recorded electron signal intensities, based on SAM and OAM incident light, at select wavelengths from 690 to 810 nm (in increments of 5 nm). Input light in the form of right and left circular polarization (labelled as $\ket{R}$ and $\ket{L}$), horizontal linear polarization and q-plates (labelled as q= -1/2, +1/2, +1) and circular polarization and a q-plate (labeled as $\ket{L,-2}$ and $\ket{R,+2}$) was detected and the singal was recorded using the photonic device at different wavelengths from 690 to 810 nm. The graphs show similar spectral shapes with peak maxima at a wavelength of 710 nm and minima at 810 nm.The decrease in the shape of the curve is due the surface electron affinity of the GaAs photonic device over time (decrease in quantum efficiency).
\begin{figure}[h]
\includegraphics [width=80mm, height=60mm] {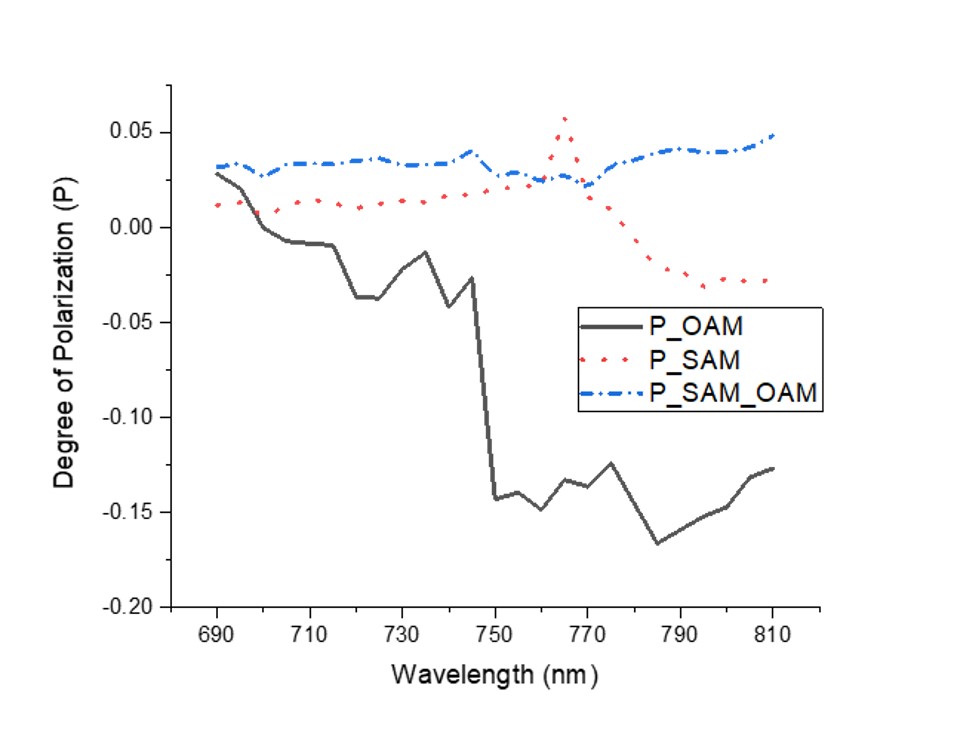}
\caption{Degree of polarization of the photogenerated electron signal pumped with SAM and OAM incident light beams versus wavelength (nm)}
\end{figure}

The degree of polarization (P) of the photogenerated electron signals was assessed for each of the three different cases of optical pumping by using the recorded photogenerated electron signal intensity (I) peaks from the spectra illustrated in Fig. 4. The degree of polarization ($P_{SAM}$) of the electron was calculated using the photogenerated electron signal pumped with $\ket{L}$ and $\ket{R}$ incident light and using equation (18)
\begin{equation}
P_{SAM} =\frac{I_{\ket{R}}-I_{\ket{L}}}{I_{\ket{R}}+I_{\ket{L}}},
\end{equation}
where $\ket{R}$ and $\ket{L}$ represent the right and light circularly polarized beam (SAM). Similarly, the degree of polarization ($P_{OAM}$) of the electron signal pumped with OAM (q=+1/2 and q=$-1/2$) incident light was calculated using equation (19)
\begin{equation}
P_{OAM} =\frac{I_{\ket{q=+1/2}}-I_{\ket{q=-1/2}}}{I_{\ket{q=+1/2}}+I_{\ket{q=-1/2}}},
\end{equation}
Lastly, the degree of polarization ${(P}_{SAM, OAM\ (q=1)})$ of the electron signal pumped with $\ket{R}$ and $\ket{L}$ and q=+1 incident light was calculated using equation (20)
\begin{equation}
P_{SAM, OAM (q=1)} =\frac{I_{\ket{R, q=+1}}-I_{\ket{L, q=-1}}}{I_{\ket{R, q=+1}}+I_{\ket{L, q=-1}}},
\end{equation}

When the photon energy $h\nu$ is greater than $(E_g +\Delta)$, above $h\nu$ $\sim$ 1.77 $eV$, electrons from the so valence band and the upper bands will be excited simultaneously. This causes the polarization to approach zero due to equal numbers of photoexcited electrons with spins up and down [5, 18, 33]. At an excitation wavelength of 695 nm $(E=1.78$ $eV$), above the bandgap, the degree of polarization was found to be $P_{OAM}=~2.1 \%$, $P_{SAM}=~1.3 \%$  and $P_{SAM, OAM (q=1)}= ~3.4\%$. These results are compared with the energy at $E=1.55$ $eV$, near bandgap, $P_{OAM}=~-14.7\%$, $P_{SAM}=~-2.6\%$  and $P_{SAM, OAM (q=+1)} = ~4.0\%$. Overall, $P_{SAM, OAM}$ shows little change throughout the wavelength spectrum, remaining at $\sim 2.5 \%$, which is close to null. The greatest variation in polarization was from $P_{OAM}$, with a prominent peak at a wavelength of 790 nm equal to $-15.9021\%$ for OAM $ l=+1$. As $h\nu < (E_g +\Delta)$, $P_{OAM}$ becomes negative with decreasing energy. The negative sign of P is attributed to the spin orientation being opposite to the angular momentum of the incident photon and may be due to the electrons with transitions from the split-off band with a negative orientation [18, 34]. 

Limited studies have been reported on the use of OAM to optically pump electrons in GaAs. The transfer of SAM and OAM beams to electron spins in GaAs was investigated using a GaAs photocathode photonic device. The degree of polarization the electron varied due to the wavelength of the excitation beam, the handedness of the beam and the value of $l$. The use of azimuthal OAM light, without the radial component, was observed. For so transitions, $P_{OAM}=~2.1 \%$, while $P_{SAM}=~1.3 \%$  and $P_{SAM,OAM (q=+1)}= ~3.4\%$. For near bandgap transitions $P_{OAM}=~-14.7\%$, $P_{SAM}=~-2.6\%$  and $P_{SAM,OAM (q=+1)} = ~4.0\%$. These results suggest that light carrying OAM, generated using q-plate of different topological charge, may be utilized to transfer momentum efficiently to the spin-polarized electrons in GaAs, and may ultimately be used in the field of spintronics. Our future work will focus on the spin relaxation and recombination times of the spin-polarized electrons using SAM and OAM beams. 

We thank Dr. Dan Nolan from Corning Inc. and Yury Budansky from IUSL for their help. We thank Earl Hergert, Dr. Ken Kaufmann and Stephanie Butron from Hamamatsu Photonics Corperation for providing information on the samples. We also thank Dr. Andrei Afanasev and Maria Solyanik for their lively discussion. We thank Professor Lorenzo Marrucci of the Universita' di Napoli Federico II for supplying q-plates. We thank Corning Inc. for their partial support. Laura A. Sordillo acknowledges financial support from Corning Inc. through the 2017-2018 and 2018-2019 Corning PhD Fellowship Awards.

\end{document}